# Monte Carlo study of the magnetic properties of the spinel $ZnFe_2O_4$ compound

**R. KHALLADI[1], S. IDRISSI[1], N. EL MEKKAOUI[1], H. LABRIM[2], I. EL HOUSNI[1], S. ZITI[3], S. MTOUGUI[1] and L. BAHMAD[1,*]**

[1] Laboratoire de la Matière Condensée et des Sciences Interdisciplinaires (LaMCScI), Mohammed V University of Rabat, Faculty of Sciences, B.P. 1014 Rabat, Morocco.

[2] USM/DERS/Centre National de l'Energie, des Sciences et des Techniques Nucléaires (CNESTEN), Rabat, Morocco.

[3] Intelligent Processing and Security of Systems, Mohammed V University of Rabat, Faculty of Sciences, B.P. 1014 Rabat, Morocco.

## Abstract

In this work, a study of the magnetic behavior of the spinel $ZnFe_2O_4$ is presented by using the Monte Carlo simulations (MCS). The iron atoms provide the magnetism in this material. In fact, the magnetic spin of moment of the $Fe^{3+}$ ions is S=5/2. In this context, we propose a Hamiltonian describing and modeling this compound. Firstly, at zero temperature we have studied the ground state phase diagrams of the system in order to find the more stable configurations. On the other hand, for a non-null temperature values, we obtained the results of the Monte Carlo simulations, namely: the magnetizations and the susceptibilities as a function of temperature. The behavior of those physical parameters shows an anti-ferromagnetic behavior with a Neel-temperature of about $T_N \approx 12$ K. This value is found to be in good agreement with the literature. Moreover, the effect of varying the crystal field, the exchange coupling interactions and the external magnetic field is studied and discussed on the behavior of the magnetizations. To complete this study, we presented the magnetic hysteresis loops deducing the antiferromagnetic behavior of the $ZnFe_2O_4$ compound.

---

[*]) Corresponding author: bahmad@fsr.ac.ma (B.L.); khalladirajaa17@gmail.com (R. K.)

# I. Introduction

Nowadays, the contribution of magnetic spinel materials have become indispensable in modern technology. Such materials are used in many electromechanical and electronic devices. Spinel ferrites MFe2O4 (MFO) type is the most studied magnetic material where the metal-iron ratio controls the magnetism of MFO nanoparticles [1]. This family of compounds has the advantage of existing in solid form, nanoparticles, thin layer and even core-shell form and gives it the opportunity to be used in various fields [2]. Also, these material types have several interesting qualities, like superior magnetic properties, corrosion resistivity, low cost and instance good chemical stability [3, 4]. In literature, this type of compound is effective and functional in photocatalysts, organic reaction catalyst fields because of the gap band that it contains [5], gas sensors [6] very sensitive magnetic resonance imaging (MRI) contrast agents [7] biomedicine [8] and radar absorbing materials [9]. Generally spinel ferrites crystallize in the cubic MgAl2O4 type structure. They have the general formula $AB_2O_4$, where B corresponds to a trivalent cation (Fe for the case of ferrites) and A, a divalent cation (Co, Zn, Ni, Mn, Sn, Fe, ...), they belong to the space group Fd3m (N ° 227 in the international tables) [10]. In addition, typical spinel MFe2O4 has antiferromagnetic properties below the Néel temperature, while at room temperature it becomes ferromagnetically [11, 12]. For the zinc ferrites ZnFe2O4 (ZFO), the Néel temperature is below 10 K [13]. Whereas a higher Néel temperature of 43 K is exhibited for epitaxial ZnFe2O4 thin films [14]. On the other hand, Stewart *et al.* [15] used X-Ray absorption near edge spectroscopy (XANES), Mossbauer spectroscopy (EM) and magnetic circular dichroism (XMCD) to show the behavior of the ZFO nanoparticles that present cationic inversion at room temperature and they found it to be ferrimagnetic. Moreover, a magnetic and structural study of Fe implanted ZnO was presented by Sheng qiang Zhou *et al.* [16] where they found that the implantation of $Fe^+$ with high fluence form a superparamagnetic α-Fe nanoparticles. Thus, many studies have found that the passage of this type of spinel from bulk to nano-size shows a modification of the long-range magnetic order which is due to the distribution of non-equilibrium cations between the tetrahedral sites (A) and octahedral (B). Also, an improvement in the magnetic response of the compound will be made [17-21]. Occupancy of both A and B sites by Iron III in ZnFe2O4 has been widely demonstrated by X-ray absorption [22–25], Mössbauer spectroscopy [26–28], magnetic measurements [29, 30].

In this work and also in one of our recent works, we have used the VESTA software [31], in order to illustrate the geometry of the studied compound, and applied the Monte Carlo simulations and the mean field theory, to establish the magnetic properties of an intermetallic compound [32]. Also, the Monte Carlo method and Ab-initio calculations were performed to predict and discuss the electronic, magnetic and the critical behavior of different compound used in spintronic applications. Such methods have been used in some of our recent works [33, 34].

The aim of this work is to investigate the magnetic properties of the spinel ferrites $ZnFe_2O_4$ compound by using Monte Carlo Simulations under the Metropolis algorithm. This manuscript will be presented as follow: firstly the ground state phase diagram of the studied system is established. Secondly we discussed the obtained results of the MCS method. Finally, a conclusion summarizing the work is given.

## II. Model and Hamiltonian

Ferrites groups (Zn ferrite) crystallize in the spinel face centered cubic (FCC) structure (see Fig.1, plotted using Vesta software [31]). These compounds are characterized by an atomic arrangement of two sites for the cations: sites A (tetrahedral oxygen coordination) and sites B (octahedral oxygen coordination). In fact, these compounds are magnetic and this magnetism comes from the Fe atoms. The magnetic spin moment of such compounds is described by S=5/2 Ising model. In order to study the magnetic properties of the specific compound $ZnFe_2O_4$, we propose the following Hamiltonian:

$$\mathcal{H} = -J_1 \sum_{<i,j>} S_i S_j - J_2 \sum_{<k,l>} S_k S_l - J_3 \sum_{<m,n>} S_m S_n - H \sum_i S_i - \Delta \sum_i S_i^2 \quad (1)$$

Where, the first, second and the third summations run for the first, second and third nearest-neighbor sites of the Fe atoms denoted by $<i,j>$, $<k,l>$ and $<m,n>$ corresponding to the exchange coupling interactions J1, J2 and J3, respectively. Due to the geometry of the system, the constraint $J_2 = J_1\sqrt{2}$ is imposed. The magnetic spin moments are: Si= ± 5/2; ± 3/2; ± 1/2. Δ stands for the crystal field acting on the Fe atoms and H is the external magnetic field applied on the all spin of the system

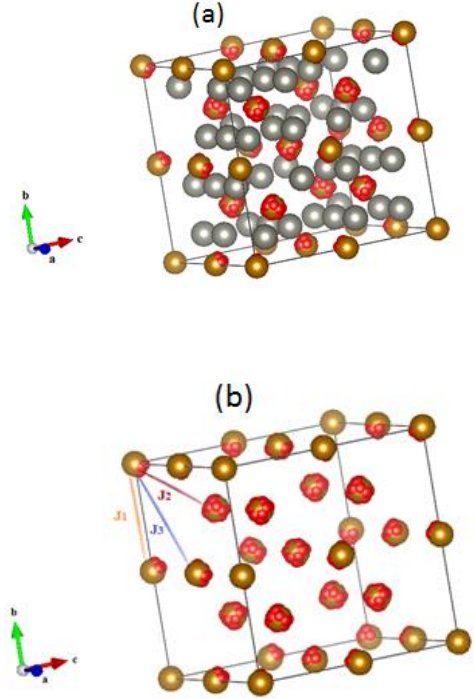

Fig.1: A sketch of the geometry of the compound ZnFe2O4 plotted using the VESTA Software [31]. In (a) the face centered cubic (FCC) structure with all atoms; in (b) the different exchange coupling interactions J1, J2 and J3.

## III.  Monte Carlo simulation

In order to compute the energies of the stable configurations, we use the Hamiltonian in Eq. (1) of the studied system. For this purpose, we perform the Monte Carlo (MC) simulation under the Metropolis algorithm. We generate several configurations, when running the MC program to reach the equilibrium. We make single-spin flip attempts, to accept or reject these changes. We apply the free boundary conditions; starting from different configurations we generate a number of $10^5$ Monte Carlo steps, discarding the first $10^4$ generated ones. In fact, we average only over the configurations reached after the discarding ones. For each parameter, we perform the Monte Carlo simulations, averaging over many initial conditions. To finalize our calculations, we apply the well-known Jackknife method [35]. Consequently, the energy and total magnetization, the magnetic susceptibility and the specific heat of the system are calculated with the following expressions:

The total energy of the system per site is:

$$\boldsymbol{E_{tot}} = \frac{1}{N} < \mathcal{H} > \quad (2)$$

Where N stands for the total number of Fe atoms in the supercell unit.

The total magnetization is:

$$M = \frac{1}{N} < \sum_i S_i > \quad (3)$$

The total susceptibility is:

$$\chi = \frac{\beta}{N}(\langle M^2 \rangle - \langle M \rangle^2) \quad (4)$$

The total specific heat of the system is given by:

$$c_v = \frac{\beta^2}{N}(\langle E_{tot}^2 \rangle - \langle E_{tot} \rangle^2) \quad (5)$$

With: $\beta = \frac{1}{K_B T}$, where T is the absolute temperature, $K_B$ is the Boltzmann constant fixed at its unit value ($K_B=1$). Hence, in all this work the physical parameters are given in their (MKSA) units.

## III.1 Ground state phase diagrams

Using the equation of the Hamiltonian, see Eq. (1), we compute different energies corresponding to different configurations. The most stable configuration corresponds to the minimum value of the energy given by this Hamiltonian. In order to study the stable configurations in the plane (H, D) for J1=J3=1, we plot in Fig.2.a the obtained stable configurations. From this figure it is found that the all possible 2.S+1=6 (for the spin moment S=5/2) configurations are found to be stable. All the configurations coexist in the point (H=0, D=-4.69), also for D<-4.69 the all possible phases are present in this figure. For D>-4.69, only two phases are found to be stable namely: the phase +5/2 for H<0 and the phase -5/2 for H>0. This is because of the antiferromagnetic behavior of the ZFO compound. In the plane (H, J1), illustrated in Fig.2.b for J3 =1 and D=1, the all possible configuration are still present in this figure, the only difference is that the area occupied by the phases +3/2 and -3/2 has been expanded. To inspect the stable configurations in the plane (D, J1), we plot in Fig.3.a the stable configurations for J3=1 and H=0. The only stable configurations in this figure are ±5/2 and ±1/2. From Fig.3.a, when fixing the value of the crystal field D, all the possible stable phases can be obtained by varying the values of the exchange coupling interaction J1 and reciprocally. Concerning the stable phases found in the plane (D, J3) for J1 =1 and H=0, we provide in Fig.3.b the obtained results. From this figure, it is found that for the point (D=-4, J3=1) the all stable phases ±1/2, ±3/2 and ±5/2 coexist. It is found that for J3<1 the stable phases are 5/2 and ±1/2 and for J3>1 the stable ones are -5/2 and ±3/2 when varying the crystal field D. On the other hand, when plotting the stable phases in the plane (J3, J1), see Fig.4, for D=1 and H=0 the only stable phases are -5/2, -3/2 and -1/2. In particular for J3=0, when J1<-2.8 the stable phases are -3/2 and -5/2; while for J1>-2.8 the only stable phase is -1/2.

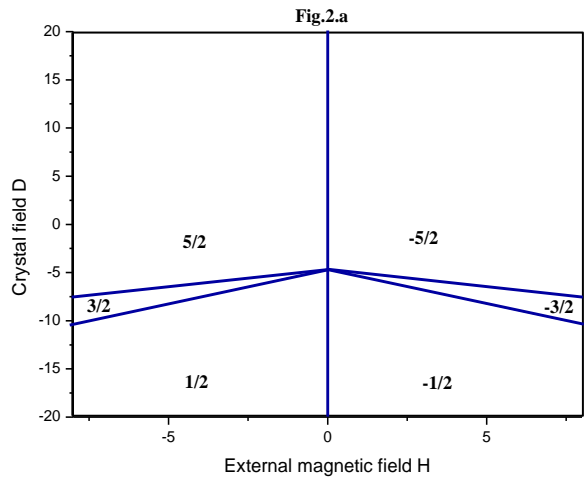

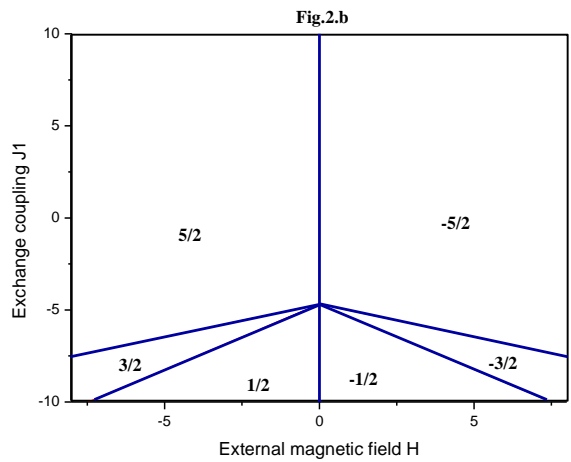

**Fig.2.** Ground state phase diagram in the plane (H, D) for $J_1=J_3=1$ **(a)** and in the plane (H, $J_1$) for $J3 =1$ and D=1 **(b).**

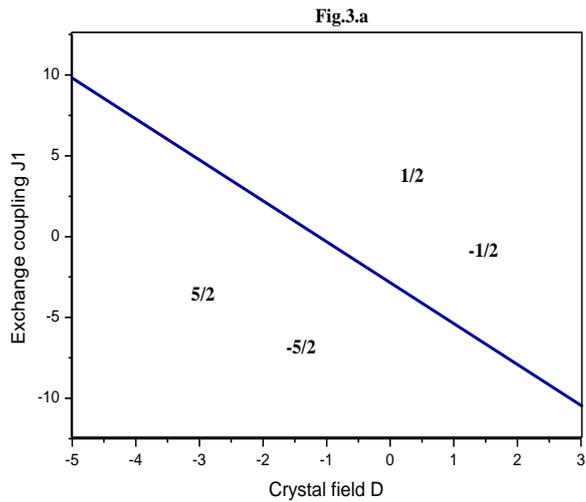

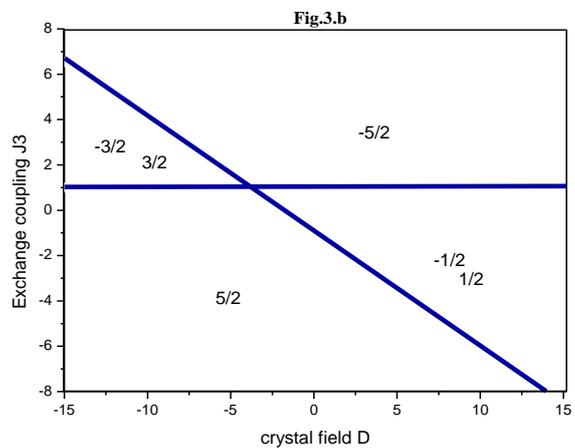

**Fig.3.** Ground state phase diagram in the plane (D, J1) for $J_3=1$ and H=0 **(a)** and in the plane (D, $J_3$) for J1 =1 and H=0 **(b).**

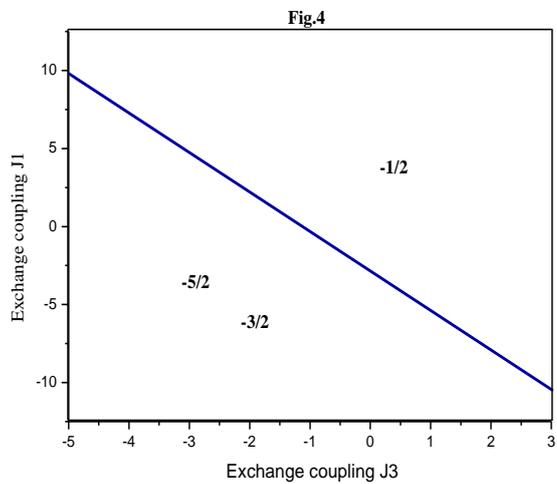

**Fig.4.** Ground state phase diagram in the plane (J3, J1) for D=1 and H=0.

## III.2 Monte Carlo results

In order to inspect the critical behavior of the spinel ZnFe2O4 we use the Monte Carlo simulations under the Metropolis algorithm. We apply the free boundary conditions when sweeping different sites of the supercell unit 3×3×3. We perform our calculations for different configurations. Starting from different initial configurations we generate $10^5$ Monte Carlo steps, discarding the first $10^4$ ones. We average only over the configurations that reached the equilibrium.

In Fig.5 we plot the behavior of the total magnetization and magnetic susceptibility of the spinel ZnFe2O4 in the absence of the external magnetic field H = 0, and fixed values of the crystal field D =1and the exchange coupling interactions: J1=1 and J3=2. From this figure it is seen that for very low temperature values the total magnetization reaches the value 2.5 which is confirmed by the ground state study. On the other hand, the behavior of the susceptibility presents a peak around the value of the transition temperature which is about $T_N \approx$22 K. It is well known that this value depends on both the crystal field and the exchange coupling interaction values.

When varying the crystal field, Figs.6.a and 6.b show the thermal behavior of the total magnetizations and specific heat for H = 0, J1 =1, J3= 2 and for selected values of the crystal field D = 0 and 2. From figure 6.b it is found that the effect of increasing the crystal field is to increase the Neel temperature from $T_N \approx$12 K, for D=0 to $T_N \approx$16 K for D=2. Also, the effect of increasing crystal field is to increase the specific heat peak amplitude, see Fig.6.b.

When exploring the effect of varying the crystal field on the behavior of the total magnetization, we provide in Figs.7.a and 7.b, the obtained results for J1=1 and J3=2.In fact, Fig.7.a is plotted for different temperature values T=10, 20 and 40 K in the absence of the external magnetic field (H=0). From this figure it is found that the paramagnetic phase is dominant for D<-7. While for D>-7 the ordered phase appears and persists for positive and large values of the crystal field. The temperature effect is negligible except for higher values of the crystal field, see Fig.7.a.

The effect of varying the external magnetic field is shown in Fig.7.b for a fixed temperature (T=10 K) and the selected values of the external magnetic field H=-1, 0 and 1. It is found that for crystal field values less than -7, the system undergoes the paramagnetic phase. On the other hand, for D>-7 the magnetization of the system undergoes the saturation value following the sign of the external magnetic field H. Also from this figure it is seen that the effect of the external magnetic field weakened for D<-7.

To show the effect of varying different exchange coupling interactions on the behavior of the total magnetizations, we give in Figs.8.a, 8.b, 8.c and 8.d the obtained results for H=0.

Indeed, Fig.8.a illustrate the effect of varying the exchange coupling interaction J1 for a fixed temperature value (T=10 K) and different crystal field values: D=-2, 0 and 1. It is seen that for J1 taking negative values the paramagnetic phase is the one that exists. While for positive value of this parameter, the total magnetization undergoes it saturation respecting the sign of the crystal field. In fact, this saturation reaches the value +2.5 for D=1, while this saturation value is -2.5 for D=-2 or 0, see Fig.8.a.

To inspect the temperature variation on the behavior of the total magnetizations, we provide in Fig.8.b our findings for D=1 and selected values of temperature: T= 10, 30 and 40 K. A first transition is appearing for J1=0 from the paramagnetic phase to the ordered one. The increasing temperature effect is to delay the saturation of the total magnetization and vice versa.

The same behavior of the total magnetization as in Fig.8.b is found in Fig.8.c when varying the exchange coupling J3 for D=1 and the selected values of temperature: T= 10, 30 and 50 K. the first order transition from the paramagnetic phase to the ordered phase is present for J3=0.

To complete the study of varying the exchange coupling interactions on the behavior of the total magnetizations we present in Fig.8.d such behavior when varying the parameter J2, for the selected values of the crystal field: D=-1, 0 and 2. The same findings are still present in this figure according to behavior of the total magnetizations. In fact, the magnetizations transit from the paramagnetic phase to the ordered one respecting the sign of the crystal field D, see Fig.8.d.

In order to complete this study, we provide in Figs.9.a, 9.b and 9.c, the obtained hysteresis loops for the compound ZnFe2O4.

In fact, Fig.9.a is plotted for fixed values of the parameters J1=1, J3=2, D=1 and different the temperature values: T= 10 and 20 K. From this figure it is seen that the surface of the hysteresis loop decreases when the temperature increases from T=10 K to T=20 K. Such result is expected according the earlier studies of our works [33, 34, 36, 37].

A reverse phenomenon is shown in Fig.9.b when varying the crystal field. In fact, the surface of the hysteresis cycle increases when the crystal field increases from D=0 to D=2.

In Fig.9.c we illustrate the effect of varying the exchange coupling interactions on the hysteresis cycles for a fixed crystal field value (D=1) and different values of the exchange coupling interactions (J1=1, J3=-1) and (J1=1, J3=2). From this figure it is seen that the surface of the hysteresis loops decreases sharply when the exchange coupling J3 takes a negative value, see Fig.9.c. This is due to the antiferromagnetic behavior of the compound ZnFe2O4 for negative values of the exchange coupling J3.

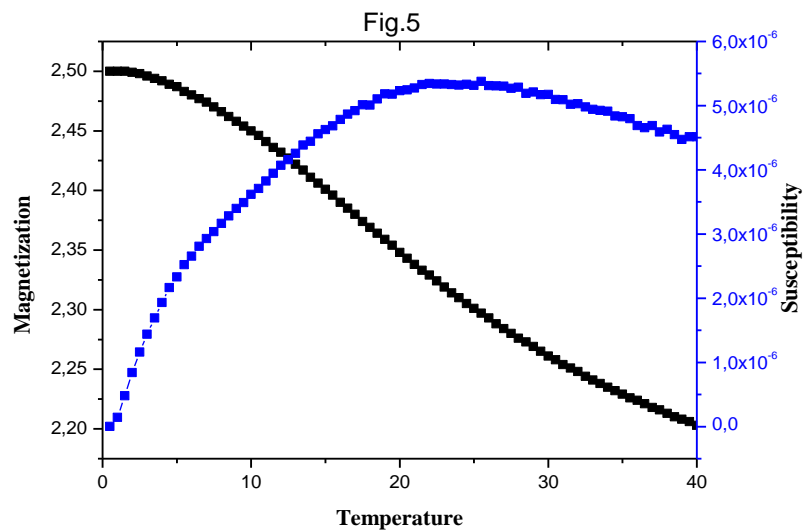

Fig.5. The thermal behavior of the total magnetization and magnetic susceptibility for the spinel ZnFe2O4 with the fixed values: H = 0, D =1, J1=1 and J3=2 .

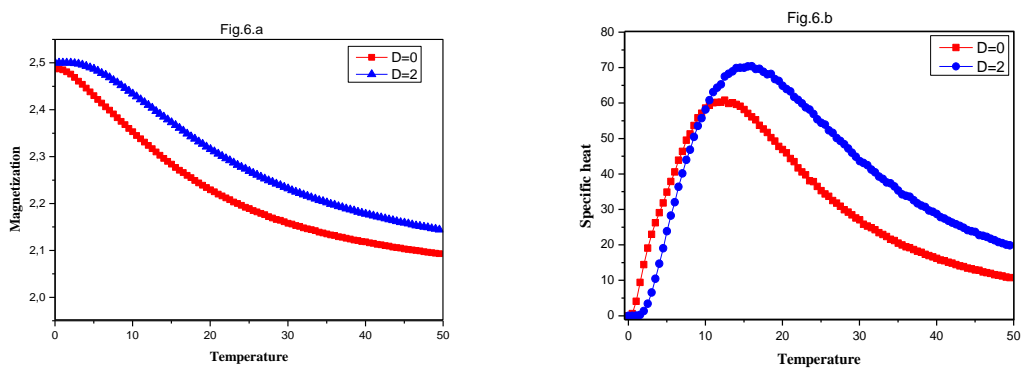

Fig.6.The thermal behavior of the total magnetizations (a) and specific heat (b) for H = 0, J1 =1, J3= 2 and for selected values of the crystal field D = 0 and 2.

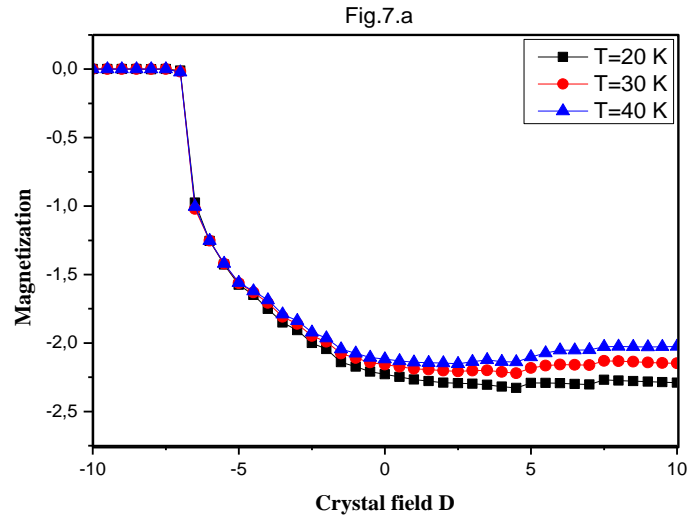

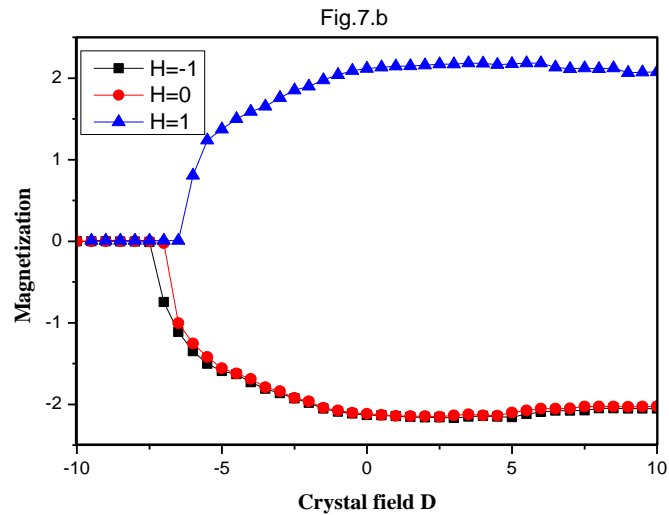

Fig.7.The total magnetization as a function of the crystal field for J1=1 and J3=2 in (a) for different temperature values T=10, 20 and 40 K and for H=0, in (b) for a fixed temperature (T=10 K) and different external magnetic field values H=-1, 0 and 1.

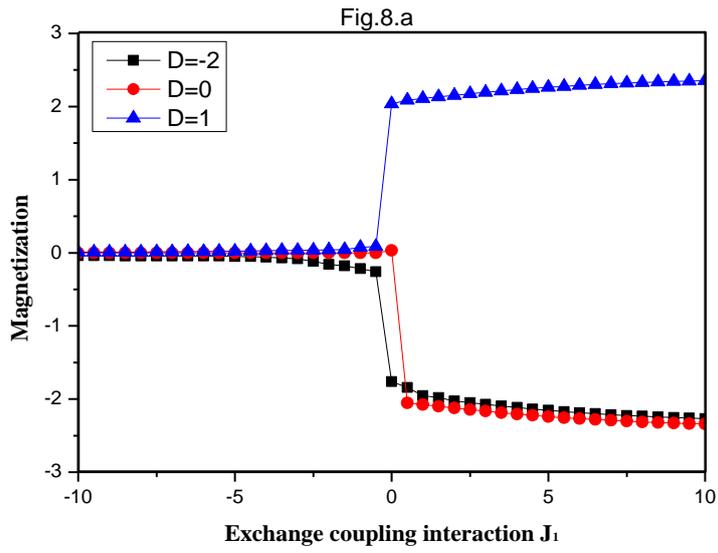

Fig.8.a

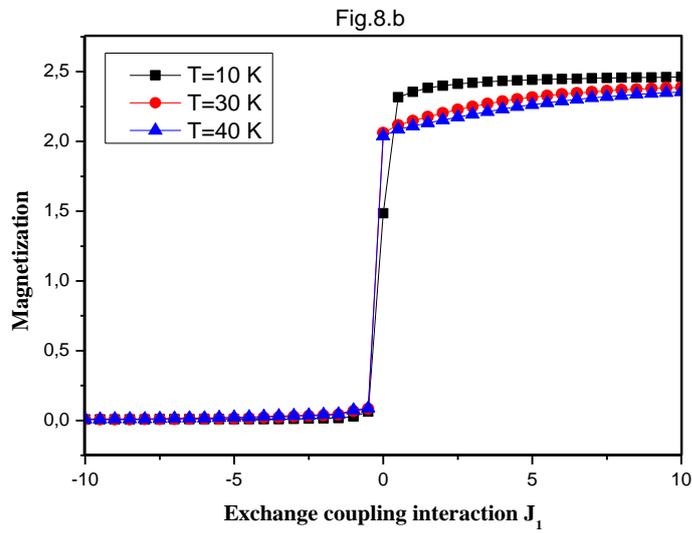

Fig.8.b

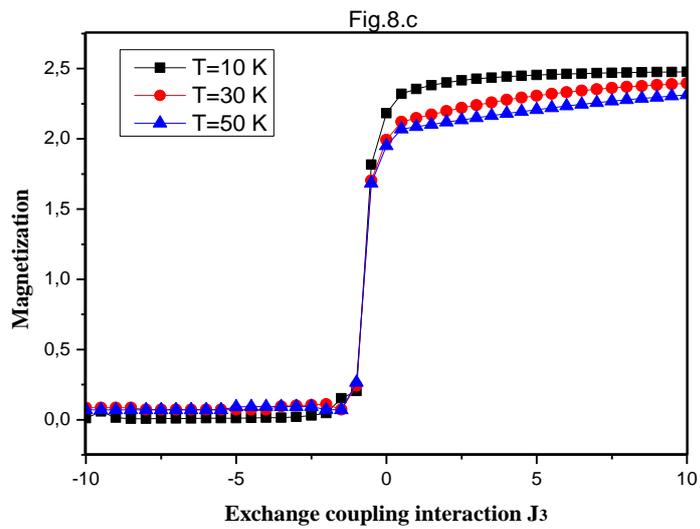

Fig.8.c

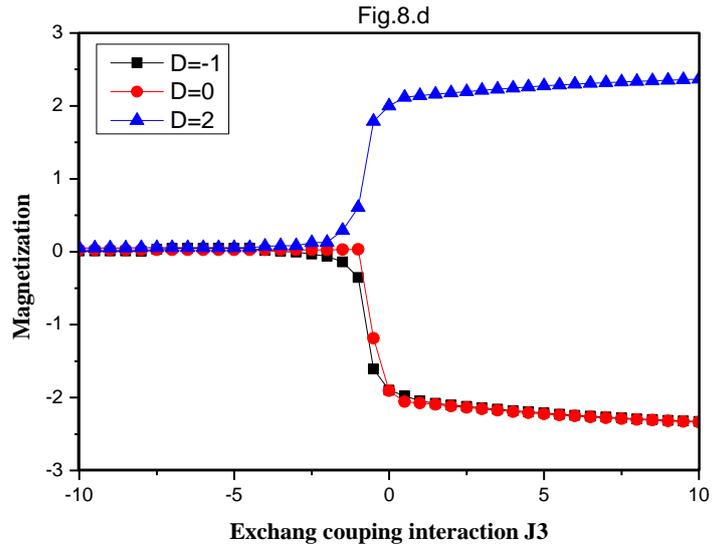

Fig.8 .The total Magnetization for H=0, as a function of the exchange coupling interactions: in (a) J1 for different crystal field values D=-2, 0 and 1 and T=10 K, in (b) J1 for selected values of temperature T= 10, 30 and 40 K and D=1, in (c) J3 for selected values of temperature T= 10, 30 and 50 K and D=1, in (d) for different values of crystal field D=-1, 0 and 2.

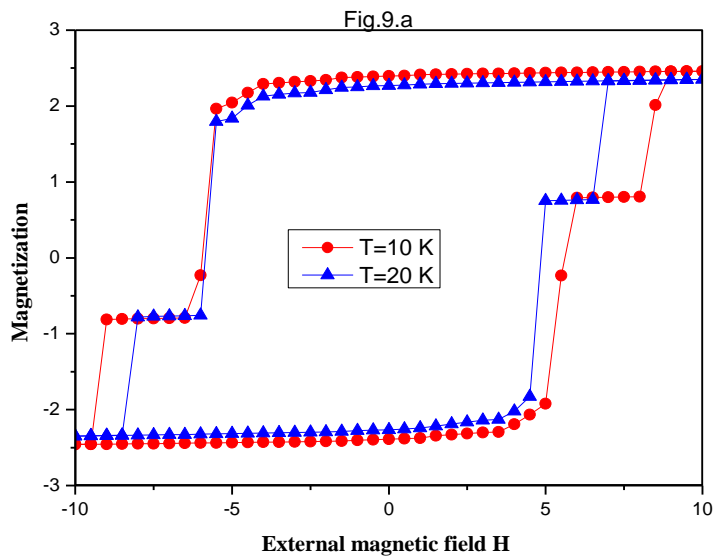

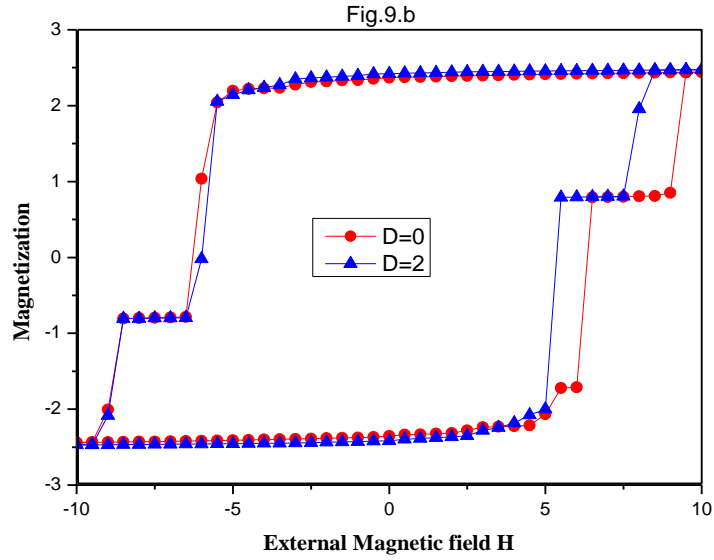

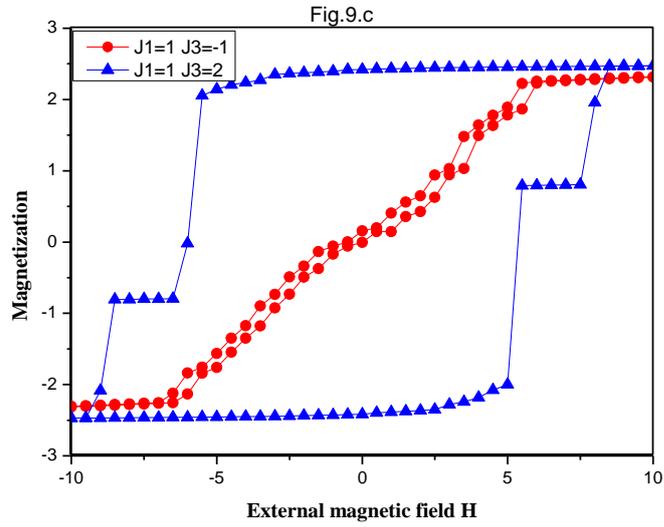

**Fig.9** The hysteresis loops for the ZnFe2O4 compound (a) for J1=1, J3=2, D=1 and selected values of the temperature T= 10 and 20 K, (b) for T=10 K and variable values of the crystal field D= 0 and 2 (c) for D=1 and different values of the exchange coupling $J_1$ and $J_3$ (J1=1, J3=-1) and (J1=1, J3=2).

## IV. Conclusion

A study of the magnetic behavior of the spinel $ZnFe_2O_4$ is presented in this paper by using the Monte Carlo simulations (MCS). For this purpose, we have proposed a Hamiltonian describing and modeling this system. In a first study, we have provided the ground state phase diagrams of the system showing the more stable configurations. In a second step, we have performed a Monte Carlo study to simulate the magnetizations and the susceptibilities as a function of temperature. An anti-ferromagnetic behavior with a Neel-temperature of about $T_N \approx 12$ K is obtained. This value is found to be in good agreement with the existing literature. On the other hand, we have presented the effect of varying the crystal field, the exchange coupling interactions and the external magnetic field on the behavior of the total magnetizations. Finally, we have presented and discussed the magnetic hysteresis loops of the $ZnFe_2O_4$ compound as a function of the external magnetic field for fixed values of the other physical parameters.